\documentclass[11pt,twoside]{article}
\usepackage{geometry}
\usepackage[T1]{fontenc}        

\usepackage[utf8x]{inputenc}    
\usepackage{ucs}                
\usepackage[numbers,super]{natbib}
\usepackage{dsfont}
\usepackage{amsmath}            
\usepackage{amsfonts}
\usepackage{amssymb}
\usepackage{booktabs}
\usepackage{dcolumn}
\usepackage{amsthm} 
\usepackage{ifthen}
\usepackage{color}
\usepackage{graphicx} 
\usepackage{tabularx}
\usepackage{textcomp}            
\usepackage{palatino}            
\usepackage{eulervm}            
\usepackage{varioref}            
\usepackage[colorlinks=false,plainpages=false]{hyperref}
\usepackage{url}                
\usepackage{cleveref}             
\usepackage{fancyhdr}
\usepackage{float}
\usepackage{multirow}
\usepackage{rotating}
\usepackage{subfig}
\usepackage{xr} 
\usepackage{xcite}
\usepackage{enumerate}

\begin{document}

\title{\Large Estimating and comparing adverse event probabilities in the presence of varying follow-up times and competing events}
\author{Regina Stegherr$^{1,*}$, Claudia Schmoor$^{2}$, Michael L{\"u}bbert$^{3}$, Tim Friede$^{4}$ \\and Jan Beyersmann$^{1}$\\[1ex] 
}
\date{January 07, 2020}
\maketitle
\noindent${}^{1}$ {Institute of Statistics, Ulm University, Ulm, Germany}\\
\noindent${}^{2}$ {Clinical Trials Unit, Faculty of Medicine and Medical Center - University of Freiburg, Freiburg, Germany}\\
\noindent${}^{3}$ {Hematology, Oncology, and Stem-Cell Transplantation, Faculty of Medicine and Medical Center - University of Freiburg, Freiburg, Germany}\\
\noindent${}^{4}$ {Institut f\"ur Medizinische Statistik, Universit\"atsmedizin G\"ottingen, G\"ottingen, Germany}

\begin{abstract}\noindent
Safety analyses in terms of adverse events (AEs) are an important aspect of benefit-risk assessments of therapies. Compared to efficacy analyses AE analyses are often rather simplistic. The probability of an AE of a specific type is typically estimated by the incidence proportion, sometimes the incidence density or the Kaplan-Meier estimator are proposed. But these analyses either do not account for censoring, rely on a too restrictive parametric model, or ignore competing events. With the non-parametric Aalen-Johansen estimator as the gold-standard, these potential sources of bias are investigated in a data example from oncology and in simulations, both in the one-sample and in the two-sample case. As the estimators may have large variances at the end of follow-up, the estimators are not only compared at the maximal event time but also at two quantiles of the observed times. To date, consequences for safety comparisons have hardly been investigated in the literature. The impact of using different estimators for group comparisons is unclear, as, for example, the ratio of two both underestimating or overestimating estimators may or may not be comparable to the ratio of the gold-standard estimator. Therefore, the ratio of the AE probabilities is also calculated based on different approaches. By simulations investigating constant and non-constant hazards, different censoring mechanisms and event frequencies, we show that ignoring competing events is more of a problem than falsely assuming constant hazards by use of the incidence density and that the choice of the AE probability estimator is crucial for group comparisons.

\end{abstract}




\noindent${}^{*}$ {Corresponding author: Regina Stegherr, e-mail: regina.stegherr@uni-ulm.de}\\
\footnotetext{\textbf{Abbreviations:} AE, adverse event, CE, competing event, RR, relative risk}

\newpage

\section{Introduction}\label{sec1}
In clinical trials, safety analyses in terms of adverse events (AEs) are a key aspect of benefit-risk assessments of therapies. Inappropriate analysis methods may result in misleading conclusions about a therapy's safety. This may lead to severe consequences for patients within the trial when the analyses were conducted for safety monitoring or for patients outside the trial when the therapy is used more widely following the trial.\cite{unkel2018} In practice, the probability of an AE of a specific type is most often estimated by the incidence proportion given by the number of patients experiencing the AE out of all patients in the respective treatment group.\cite{oneill1987} As the incidence proportion does not account for the time a patient is under observation, it ignores censoring and may lead to an underestimation of the adverse event probability.\cite{unkel2018, allignol2016} To consider time under observation, survival methods in form of the incidence density which divides by patient-time-at-risk are suggested.\cite{bender2016} In constrast to the incidence proportion, the incidence density, also called incidence rate, is that it estimates a hazard and not of a probability. The incidence density can be transformed to the probability scale as we will see later. This could be interpreted as a parametric version of one minus the non-parametric Kaplan-Meier estimator, but imposing the assumption of constant hazards over time, which is a typical point of criticism.\cite{kraemer2009} 

Additionally to censoring, competing events (CEs) such as death or premature treatment-related discontinuation of participation in the study can occur preventing the observation of the AE.\cite{allignol2016} The Kaplan-Meier estimator and the probability transform of the incidence density treat the CEs as censored observations and therefore do not account for CEs. As a result, they overestimate the AE probability.\citep{lacny2015, lacny2018}

An estimator accounting for all three potential sources of bias, namely censoring, non-constant hazards and CEs, is the non-parametric Aalen-Johansen estimator.\cite{aalen1978} Therefore, it is considered the gold standard. Under the assumption of constant hazards for the AE and for the CE, a parametric version of the Aalen-Johansen estimator of the AE probability can be constructed from these two hazards. This estimator is called the probability transform of the incidence density accounting for CEs. 

The problems related to biased estimation of the AE probability have been previously described.\cite{unkel2018,allignol2016,bender2016,hollaender2019} Here we extend the discussion by investigating the following three questions: (i) What is the impact of choosing different estimators of the AE probabilities on group comparisons in terms of bias and precision? (ii) Is the impact of ignoring CEs possibly worse than falsely assuming constant hazards? (iii) How does the time point of analysis influence the previous two questions, especially for the incidence proportion?

Regarding question (i), one aspect of the benefit-risk assessment of safety analyses is treatment comparisons in terms of the relative risk. The relative risk compares two treatments by taking the ratio of the estimators of the experimental treatment and the control treatment. Even if the used probability estimator may underestimate or overestimate the AE probability, the ratio of two probability estimates, obtained with one of the biased estimators, may be comparable to the ratio of the probability estimates obtained with the gold standard. In this paper additionally to the AE probability estimators also the variances of the AE probability estimators are investigated as misspecification of the hazard in the form of falsely assuming constant hazards may also influence the variances of the parametric estimators. A non-parametric bootstrap is suggested as a suitable alternative to obtain the variance estimates.\cite{hjort1992} These bootstrapped variances are compared to the asymptotic, model-based estimators to see whether the assumption of constant hazards also impacts the variances. Question (ii) aims to investigate whether a misspecified incidence density analysis that does account for CEs may be useful provided that variance estimation accounts for misspecification. 
Question (iii) is a consequence of the problem that the incidence proportion is usually only calculated at the end of follow-up in each group of the two treatment groups. This leads to different evaluation time points in the relative risk of the incidence proportion drawing the interpretation into question.\cite{kunz2015} To solve this issue all comparisons are not only conducted at the end of follow-up but also at the shorter of the two observed maximum follow-up times in the two groups. There is also the concern that a low number of observations under study at the end of follow-up leads to increased variances.\cite{pocock2002} Hence, the comparisons of the AE probability estimators are also investigated at two different quantiles of the observed times.

We also note that Bender and Beckmann\citep{bender2019} have recently investigated whether the ratio of incidence densities may serve as an estimator of the hazard ratio even under misspecification. Their investigation was also motivated by AE analyses, also considered variances (via confidence intervals) and found that results depend on the baseline cumulative AE probability. However, these authors did neither consider competing events (which impacts probabilities) nor bootstrapping variances (which may lead to larger confidence intervals).

The paper is organized as follows, Section~\ref{sec2} introduces the AE probability estimators with corresponding variance estimators. Section~\ref{sec3} presents the comparisons of the estimators at the different follow-up times based on data from an oncology trial. In Section~\ref{sec4} a simulation study addresses the three questions posed above. The paper concludes with a discussion in Section~\ref{sec5}.

\section{Estimators and their variances of event probabilities and their ratios}\label{sec2}

\subsection{Competing risks model}
\begin{figure}[t]
\centerline{\includegraphics[width=150pt,height=9pc,angle=0]{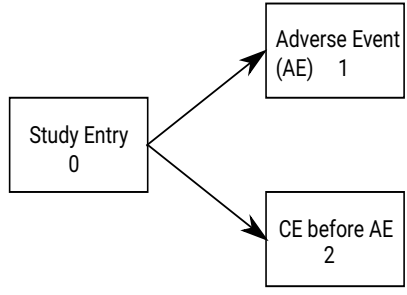}} 
\caption{Competing risks setting.\label{fig1}}
\end{figure}

In the following we consider data from a two-arm randomized controlled trial with each group following a competing risks model displayed in Figure~\ref{fig1}. Every patient starts in the initial state $0$ at study entry, i.e., at time $0$. The event time at which a patient $i$ moves from state $0$ to either state $1$ or $2$, whatever is observed first, is denoted by $T_i$ and the event type is denoted by $\epsilon_i$ as $\epsilon_i=1$ in case of an AE and $\epsilon_i=2$ in case of a CE. A patient $i$ can also be censored at time $C_i$ if $C_i<T_i$. Only the minimum of the censoring time $C_i$ or the event time $T_i$ can be observed. Therefore, the observable data consists of \textit{i.i.d.} replicates of $(\min(T_i,C_i),\mathds{1}(T_i\le C_i)\cdot \epsilon_i)$. Furthermore, $T_i$ and $C_i$ are assumed to be independent. 

In this setting the AE hazard is defined as $\lambda(t)=\lim\limits_{n\rightarrow\infty}P(T\in[t,t+\text{d}t), \epsilon=1\vert T\ge t)/\Delta t$ and the hazard of the CE as $\bar\lambda(t)=\lim\limits_{n\rightarrow\infty}P(T\in[t,t+\text{d}t), \epsilon=2\vert T\ge t)/\Delta t$ respectively.

All estimators are evaluated at time $\tau$. Later $\tau$ will take the values $\tau_{max}^{(A)}=\max\{\min(T_i,C_i)\vert i \text{ in group A}\}$ and $\tau_{max}^{(B)}=\max\{\min(T,C)\vert i \text{ in group B}\}$ which are the maximum follow-up times in the experimental group and control group, respectively. In general, $\tau_{max}^{(A)}\ne \tau_{max}^{(B)}$, but comparing both groups by evaluating estimators at the respective maximum follow-up time is what is commonly done in the analyses of adverse events. But to refrain from observing one group much longer than the other, we also consider $\tau_{max}=\min(\tau_{max}^{(A)},\tau_{max}^{(B)})$, $\tau_{P90}=\min(\tau_{P90}^{(A)},\tau_{P90}^{(B)})$, with $\tau_{P90}^{(A)}$ and $\tau_{P90}^{(B)}$ the empirical 90\% quantiles of $\min(T,C)$ in group A and B, respectively, and $\tau_{P60}$ defined in the same way.

\subsection{Estimators of event probabilities and their variances}

The following five different estimators of the AE probability are considered and will be compared.
The estimators are only displayed for the experimental group A. For the control group B, they are derived analogously.

\begin{itemize}
\item \textit{Incidence proportion}: Let $d_{\rm A}(u)$ denote the number of observed AEs at time $u$ in group A and let  $n_A$ be the total number of patients in group A, then the incidence proportion is defined as
\begin{eqnarray*}
\widehat{\text{IP}}_{\rm A}(\tau)=\frac{\sum\limits_{u\in (0,\tau]} d_{\rm A}(u)}{n_{\rm A}},
\end{eqnarray*}
where the sum is over all observed, unique event times $u$. The incidence proportion estimates $P(\text{AE in }[0,\tau],\text{AE observed})$. The corresponding model-based variance estimator is
\begin{eqnarray*}
\hat s_{\rm A}^2=\frac{\widehat{\text{IP}}_{\rm A}(\tau)(1-\widehat{\text{IP}}_{\rm A}(\tau))}{n_{\rm A}}.
\end{eqnarray*}

\item \textit{Probability transform incidence density}: We first define the incidence density as
\begin{eqnarray*}
\widehat{\text{ID}}_{\rm A}(\tau)=\frac{\sum\limits_{u\in (0,\tau]} d_{\rm A}(u)}{\sum\limits_{i=1}^{n_{\rm A}}\min(t_i,\tau)},
\end{eqnarray*}
where the denominator is the population time at risk restricted by $\tau$ with $t_i$ the time of the first event irrespective of the event type of patient $i$. The model-based variance estimator of the incidence density is
\begin{eqnarray*}
\widehat{\rm var}(\widehat{ID}_{\rm A})(\tau)=\frac{\sum\limits_{u\in (0,\tau]} d_{\rm A}(u)}{\left(\sum\limits_{i=1}^{n_{\rm A}}\min(t_i,\tau)\right)^2}.
\end{eqnarray*}
The incidence density is an estimator of the hazard $\lambda_{\rm A}$ and as a consequence not directly comparable to the incidence proportion. But by the assumption of constant hazards and the connection to the exponential distribution it can be transformed to the probability scale by $1-\exp(-\widehat{ID}_{\rm A}(\tau)\cdot \tau)$. If the assumption of constant hazards holds, $\lambda_{\rm A}(t)=\lambda_{\rm A} \forall t$, the incidence density is an unbiased estimator of the hazard. The model-based variance estimator of the probability transform is $\hat s_{\rm A}^2=\tau^2\cdot\exp(-\tau\cdot \widehat{ID}_{\rm A}(\tau))^2\cdot \widehat{\rm var} \widehat{ID}_{\rm A}(\tau)$, using the delta-method.

\item \textit{1- Kaplan-Meier}: Let $\Delta\hat\Lambda_{\rm A}(u)$ be the increment of the Nelson-Aalen estimator of the cumulative AE hazard and therefore closely related to $\widehat{ID}_{\rm A}(\tau)$, then the 1-Kaplan-Meier estimator is defined as
\begin{eqnarray*}
1-\hat{S}_{\rm A}(\tau)=1-\prod\limits_{u\in(0,\tau]} \left(1-\Delta\hat\Lambda_{\rm A}(u)\right).
\end{eqnarray*}
The Kaplan-Meier estimator estimates the event-specific survival function $S_{\rm A}(\tau)=\exp\left(-\int\limits_0^\tau \lambda_{\rm A}(u) \text{ d}u\right)$ as it treats CEs by censoring the follow-up time. Its model-based variance can be estimated using the Greenwood variance estimator.\cite{andersen1993} Note that $S_{\rm A}$ does not have a proper probability interpretation as a consequence of competing events, see below.

\item \textit{Aalen-Johansen estimator}: As gold standard we consider here the Aalen-Johansen estimator which is given by 
\begin{eqnarray*}
\widehat{AJ}_{\rm A}(\tau)=\sum\limits_{u\in(0,\tau]}\prod\limits_{v\in(0,u)} \left(1-\Delta\hat\Lambda_{\rm A}(v)-\Delta\hat{\overline\Lambda}_{\rm A}(v)\right)\Delta\hat{\Lambda}_{\rm A}(u)
\end{eqnarray*}
where $\Delta\hat{\overline\Lambda}_{\rm A}(v)$ is the increment of the Nelson-Aalen estimator of the CE. The model-based variance of the Aalen-Johansen estimator can be estimated using a Greenwood-type estimator.\cite{allignol2011} The Aalen-Johansen estimator estimates the cumulative incidence function $P(T\le \tau,\epsilon=1\vert \text{ group A})=\int\limits_0^\tau P(T\ge u-\vert \text{ group A})\lambda_{\rm A}(u)\text{ d}u$. Comparing 1 minus the event-specific survival function and the cumulative incidence function it can be easily shown than 1 minus the event-specific survival function is greater than the cumulative incidence function as long as CEs are present (see Appendix \ref{app1} for details), and the same inequality holds for the 1-Kaplan-Meier estimator and the Aalen-Johansen estimator.

\item \textit{Probability transform incidence density accounting for CEs}:
\begin{eqnarray*}
\frac{\widehat{ID}_{\rm A}(\tau)}{\widehat{ID}_{\rm A}(\tau)+\widehat{\overline{ID}}_{\rm A}(\tau)}\left(1-\exp(-\tau[\widehat{ID}_{\rm A}(\tau)+\widehat{\overline{ID}}_{\rm A}(\tau)])\right),
\end{eqnarray*}
where $\widehat{\overline{ID}}_{\rm A}(\tau)={\sum\limits_{u\in (0,\tau]} \bar{d}_{\rm A}(u)}/{\sum\limits_{i=1}^{n_{\rm A}}\min(t_i,\tau)}$ with $\bar{d}_{\rm A}(u)$ the number of observed CEs at time $u$ in group A, is the incidence density of the CE and, hence, the parametric analog of $\Delta\hat{\overline\Lambda}_{\rm A}(\tau)$. Using the incidence density of the CE, the connection between the incidence density and the incidence proportion is $\widehat{ID}_{\rm A}(\tau)/(\widehat{ID}_{\rm A}(\tau)+\widehat{\overline{ID}}_{\rm A}(\tau))=\sum\limits_{u\in (0,\tau]} d_{\rm A}(u)/n_{\rm A}$ in the absence of censoring.\cite{beyersmann2017}
 Moreover, a model-based variance estimator of the estimator of the probability transform of the incidence density accounting for CEs can be derived with the delta-method \cite{bonofiglio2016} and is provided in the Appendix \ref{app2}.
In the following, we will also call the probability transform of the incidence density accounting for CEs, somewhat loosely, parametric counterpart of the Aalen-Johansen estimator as the two estimators estimate the same quantity under the parametric assumption of constant hazards.

\end{itemize}

Another way to estimate the variances of the estimators is with a non-parametric bootstrap accounting for model misspecifications that may also influence the model-based variances.\cite{hjort1992} The variances of the parametric estimators given above assume that the hazards are constant. This assumption is not made by the non-parametric bootstrap.

\subsection{Between group comparisons}

The comparison of the two treatment groups can be done in terms of the relative risk
\begin{eqnarray*}
\widehat{RR}(\tau)=\frac{\hat p_{\rm A}(\tau)}{\hat p_{\rm B}(\tau)},
\end{eqnarray*}
where $\hat p_{\rm A}(\tau)$ is one of the above introduced estimators of the AE probability in the interval $[0,\tau]$.
The variance estimator of the relative risk can be constructed based on a log-transformation
\begin{eqnarray*}
\widehat{\rm var}(\log{\widehat{RR}}(\tau))=\frac{1}{\hat p_{\rm A}(\tau)^2}\cdot \hat s_{\rm A}(\tau)^2+\frac{1}{\hat p_{\rm B}(\tau)^2}\cdot \hat s_{\rm B}(\tau)^2 ,
\end{eqnarray*}
where $ \hat s_{\rm A}(\tau)^2$ and $ \hat s_{\rm B}(\tau)^2$ are one of the two suggested variance estimators of $\hat p_{\rm A}(\tau)$ and $\hat p_{\rm B}(\tau)$ evaluated at time $\tau$.

\section{An example: The DECIDER trial in acute myeloid leukemia}\label{sec3}
As an example we use data from the DECIDER trial (DECItabine, DEacetylase inhibition, Retinoic acid; ClinicalTrials.gov identifier: NCT00867672).\cite{grishina2015,lubbert2019} This randomized, multicenter trial had the objective to investigate the efficacy
and safety of valproate and ATRA in combination with decitabine in 200 older and nonfit patients with acute myeloid leukemia. The trial had a 2 x 2 design, in which patients were randomly assigned to 1 of 4 treatment arms: decitabine, decitabine + valproate, decitabine + ATRA, or decitabine + valproate + ATRA. Here, we consider the comparison of the combined ATRA treatment arms (called group A, $n_A=96$) with the combined no ATRA treatment arms (called group B, $n_B=104$) with respect to the adverse event severe thrombocytopenia, Common Terminology Criteria for Adverse Events (CTCAE) grade 3-5. This AE was observed in 35 patients in group A and in 32 patients in group B. Death and end of treatment plus 28 days with no observed AE during that time were considered as CEs. As ATRA prolonged overall survival time (median of 8.2 months in group A and of 5.1 months in group B\cite{lubbert2019}), a CE was experienced by 56 patients in group A and by 69 patients in group B, leading on average to a longer follow-up time for AEs in group A (mean 137 days) as compared to group B (mean 125 days).
 So in this data set there is hardly any censoring as only 5 patients in group A and 3 patients in group B were censored. The time was measured in days and analyses were performed at $\tau_{max}^{(A)}=802$, respectively $\tau_{max}^{(B)}=980$, the maximum follow-up times in the two groups, at $\tau_{max}=802$ the minimum of the two maximum follow-up times, at $\tau_{P90}=353$ and at $\tau_{P60}=67$, both chosen according to the quantiles of the observed times.

\subsection{Estimating the AE probability}\label{sec31}
\begin{figure}[t]
\centerline{\includegraphics[width=400pt,height=200pt]{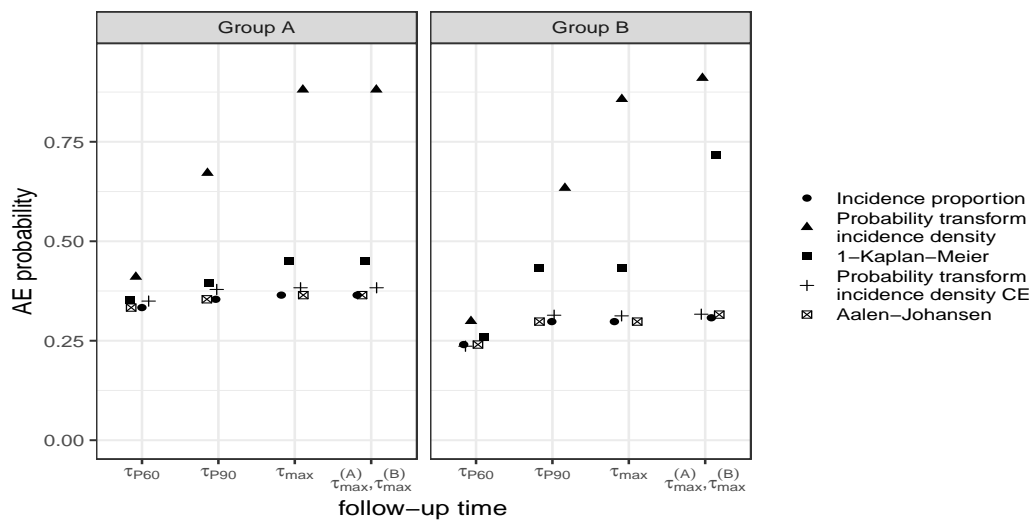}} 
\caption{Adverse event probability estimators applied to the DECIDER trial at the different follow-up times. The symbols have been jittered for better readability.\label{fig2}}
\end{figure}

\begin{figure}[t]
\centerline{\includegraphics[width=400pt,height=200pt]{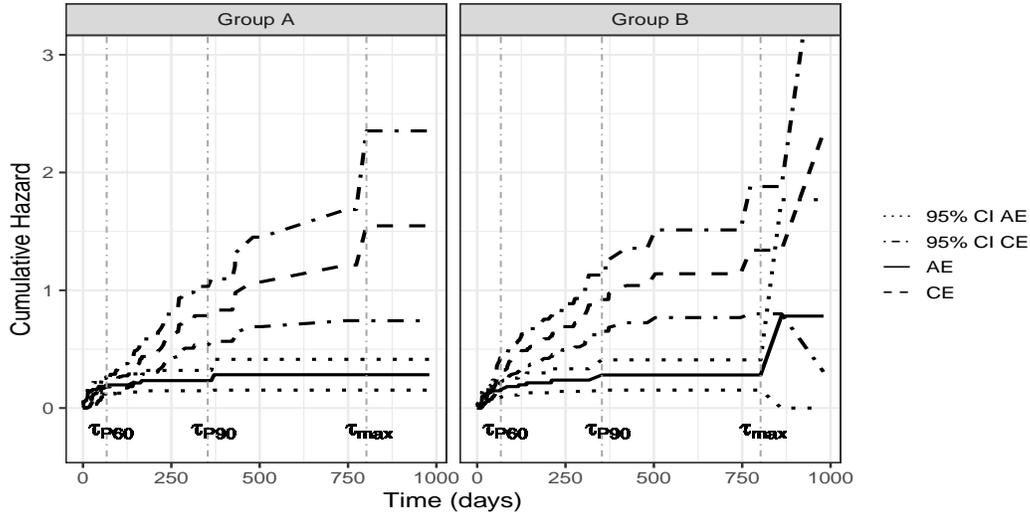}} 
\caption{Cumulative hazard and 95\% pointwise confindence intervals (95\% CI) of the adverse event and the competing event in both treatment groups. The vertical lines display the analysis times. \label{fig3}}
\end{figure}

Figure~\ref{fig2} displays the five different AE probability estimators for the four different follow-up times. In the original analysis of the trial, the AE probabilites were estimated at $\tau_{max}^{(A)}$, $\tau_{max}^{(B)}$ using the incidence proportion, leading to an estimated probability of severe thrombocytopenia of 36.5\% in group A and of 30.8\% in group B\cite{lubbert2019}. This difference was not regarded as relevant taking into account the superiority of group A with respect to overall survival. Now considering the other estimators, the probability transform of the incidence density results in a higher estimated AE probability than the Aalen-Johansen estimator at all follow-up times. The difference is more pronounced for longer follow-up times as the CEs are observed later in time. The 1-Kaplan-Meier estimator also always obtains a higher estimated AE probability than the Aalen-Johansen estimator but less pronounced than the probability transform of the incidence density. For the latter, the estimated cumulative hazards in Figure~\ref{fig3} illustrate departures from the constant hazard assumption. Moreover, Figure~\ref{fig3} shows that CEs tend to occur later in time than AEs. This is the reason why the difference between the probability transform of the incidence density or the 1-Kaplan-Meier estimator and the Aalen-Johansen estimator is larger at later follow-up times. 

The other three estimators differ only slightly. The probability transform of the incidence density accounting for CEs is comparable to the Aalen-Johansen estimator. Even though the assumption of constant hazards is not valid the difference between the two estimators that account for the CE is nearly negligible. The incidence proportion takes the same value as the Aalen-Johansen estimator as long as there are no censored observations in the data.\cite{beyersmann2017} In these data, there is hardly any censoring but the last patient under observation in group B is censored leading to a slightly lower incidence proportion compared to the Aalen-Johansen estimator. When estimating the probabilities at earlier follow-up times this difference is less pronounced as in these situations the last observation in both groups is always an AE or a CE.

\subsection{Variance estimation}\label{sec32}
The variances of the probability estimators are calculated in two ways, first by using the formulas displayed in Section~\ref{sec2} and second by a non-parametric bootstrap. 

\begin{center}
\begin{table*}[h!]%
\caption{Estimated variances of the AE probabilities using the analytically derived model-based variances and bootstrapped variances. The following follow-up times are considered $\tau_{max}^{(A)}$,$\tau_{max}^{(B)}$ the maximum follow-up times in group A and group B respectively, $\tau_{max}$ the minimum of the two maximal event times, $\tau_{P90}$ and $\tau_{P60}$ chosen according to the quantiles of the time and account for different lengths of follow-up between group A and group B.\label{tab1}}
\centering
\hspace*{-3cm}\begin{tabular*}{500pt}{@{\extracolsep\fill}clccD{.}{.}{3}c@{\extracolsep\fill}}
\toprule
& & \multicolumn{2}{@{}c@{}}{\textbf{Group A} }& \multicolumn{2}{@{}c@{}}{\textbf{Group B} }\\\cmidrule{3-4}\cmidrule{5-6}
\textbf{Follow-up time} & \textbf{Estimator} & \textbf{model-based}  & \textbf{bootstrap}  & \multicolumn{1}{@{}l@{}}{\textbf{model-based}}  & \textbf{bootstrap}   \\
\midrule
$\tau_{max}^{(A)}$,$\tau_{max}^{(B)}$&Incidence proportion & 0.0024 & 0.0025 & 0.0020 & 0.0019 \\ 
$\tau_{max}^{(A)}$, $\tau_{max}^{(B)}$&Probability transform incidence density & 0.0018 & 0.0036 & 0.0014 & 0.0034 \\
$\tau_{max}^{(A)}$, $\tau_{max}^{(B)}$&1-Kaplan-Meier & 0.0054 & 0.0060 & 0.0419 & 0.0509 \\ 
$\tau_{max}^{(A)}$, $\tau_{max}^{(B)}$& Aalen-Johansen & 0.0024 & 0.0025 & 0.0022 & 0.0020 \\ 
$\tau_{max}^{(A)}$, $\tau_{max}^{(B)}$&Probability transform incidence density CE & 0.0026 & 0.0026 & 0.0021 & 0.0020 \\ 
\cmidrule{3-6}
$\tau_{max}$ & Incidence proportion  & 0.0024 & 0.0025 & 0.0020 & 0.0019 \\ 
$\tau_{max}$  &Probability transform incidence density & 0.0018 & 0.0041 & 0.0025 & 0.0045 \\ 
$\tau_{max}$  & 1-Kaplan-Meier & 0.0054 & 0.0060 & 0.0062 & 0.0067 \\ 
$\tau_{max}$ & Aalen-Johansen & 0.0024 & 0.0025 & 0.0020 & 0.0019 \\ 
$\tau_{max}$ &Probability transform incidence density CE  & 0.0026 & 0.0027 & 0.0022 & 0.0021 \\ 
\cmidrule{3-6} 
$\tau_{P90}$ &Incidence proportion  & 0.0024 & 0.0024 & 0.0020 & 0.0019 \\ 
$\tau_{P90}$&Probability transform incidence density & 0.0039 & 0.0059 & 0.0044 & 0.0049 \\ 
$\tau_{P90}$& 1-Kaplan-Meier & 0.0031 & 0.0036 & 0.0062 & 0.0047 \\ 
$\tau_{P90}$ & Aalen-Johansen & 0.0024 & 0.0024 & 0.0020 & 0.0019 \\ 
$\tau_{P90}$ &Probability transform incidence density CE  & 0.0026 & 0.0028 & 0.0022 & 0.0021 \\ 
\cmidrule{3-6}
$\tau_{P60}$&Incidence proportion & 0.0023 & 0.0022 & 0.0018 & 0.0016 \\ 
$\tau_{P60}$ &Probability transform incidence density & 0.0030 & 0.0032 & 0.0025 & 0.0023 \\ 
$\tau_{P60}$ & 1-Kaplan-Meier & 0.0026 & 0.0024 & 0.0021 & 0.0018 \\ 
$\tau_{P60}$ & Aalen-Johansen & 0.0023 & 0.0022 & 0.0018 & 0.0016 \\ 
$\tau_{P60}$ &Probability transform incidence density CE  &0.0024 & 0.0027 & 0.0017 & 0.0017 \\ 
\bottomrule
\end{tabular*}\hspace*{-2cm}
\end{table*}
\end{center}
\vspace{-1cm}
Table~\ref{tab1} displays the estimated variances of all estimators for both groups evaluated at the different follow-up times. At the maximal follow-up time in each group, two major points can be observed. Firstly, the variance of the 1-Kaplan-Meier estimator is much larger than all other variances. This verifies the expectation that at later follow-up times the variance of the Kaplan-Meier estimator is increased due to a small number of patients still at risk and this is the motivation for comparing also at the other follow-up times.\cite{pocock2002} Also for shorter follow-up, except for $\tau_{P60}$, this increased variance can be observed. At $\tau_{P60}$ all variances are virtually the same.

Secondly, for the probability transform of the incidence density the model-based variance is smaller than the bootstrapped variance. This is likely due to the model misspecification as the constant hazard assumption might not hold for the AE hazard as Figure~\ref{fig3} suggests. Again this effect is more pronounced for longer follow-up times than for $\tau_{P60}$. 

The two variance estimators of the Aalen-Johansen estimator and of the probability transform of the incidence density accounting for CEs are almost identical as, like the incidence proportion, they aim to estimate the binomial probability $P(\text{AE in }[0,t])$.

\subsection{Group comparisons}\label{sec33}

We investigate how the ratio of two biased estimators of the probabilities compares to the ratio of two unbiased estimators. Table~\ref{tab2} displays the relative risks calculated from the estimators and time points shown in Figure~\ref{fig2}. Additionally, the 95\% confidence intervals of the relative risk calculated with both variance estimators are displayed. 

\begin{center}
\begin{table*}[h!]%
\caption{Relative risk (RR), model-based confidence interval (CI), bootstrap CI and the ratio of the lengths of the two confindence intervals (model-based/bootstrap) calculated from the different AE probability estimators at several follow-up times.\label{tab2}}
\centering
\hspace*{-3.5cm}\begin{tabular*}{500pt}{@{\extracolsep\fill}clcccD{.}{.}{3}c@{\extracolsep\fill}c}
\toprule
& & &\multicolumn{2}{@{}c@{}}{\textbf{model-based} }& \multicolumn{2}{@{}c@{}}{\textbf{bootstrap} } & \textbf{ratio of}\\
& & &\multicolumn{2}{@{}c@{}}{\textbf{CI} }& \multicolumn{2}{@{}c@{}}{\textbf{CI} } &{\textbf{the lengths }}\\\cmidrule{4-5}\cmidrule{6-7} 
\textbf{Follow-up time} & \textbf{Estimator} & \textbf{RR} &\textbf{Lower}  & \textbf{Upper}  & \multicolumn{1}{@{}l@{}}{\textbf{Lower}}  & \textbf{Upper}&   {\textbf{of the CIs}}\\
\midrule
$\tau_{max}^{(A)}$, $\tau_{max}^{(B)}$ & Incidence proportion & 1.1849 & 0.8015 & 1.7517 & 0.8055 & 1.7429 &1.0138\\ 
$\tau_{max}^{(A)}$, $\tau_{max}^{(B)}$&Probability transform incidence density& 0.9678 & 0.8540 & 1.0968 & 0.8053 & 1.1631&  0.6786\\ 
$\tau_{max}^{(A)}$, $\tau_{max}^{(B)}$&1-Kaplan-Meier &   0.6280 & 0.3294 & 1.1971 & 0.3105 & 1.2703 & 0.9041\\ 
$\tau_{max}^{(A)}$, $\tau_{max}^{(B)}$& Aalen-Johansen & 1.1550 & 0.7805 & 1.7092 & 0.7843 & 1.7008 & 1.0134\\ 
$\tau_{max}^{(A)}$, $\tau_{max}^{(B)}$&Probability transform incidence density CE & 1.2097 & 0.8217 & 1.7810 & 0.8259 & 1.7719 & 1.0140 \\ 
\cmidrule{3-8}
$\tau_{max}$  & Incidence proportion &  1.2231 & 0.8233 & 1.8172 & 0.8259 & 1.8114 & 1.0085\\ 
$\tau_{max}$  & Probability transform incidence density & 1.0277 & 0.8857 & 1.1924 & 0.8337 & 1.2668 &  0.7082\\ 
$\tau_{max}$  & 1-Kaplan-Meier  & 1.0415 & 0.6449 & 1.6820 & 0.6296 & 1.7230 &  0.9485\\ 
$\tau_{max}$ & Aalen-Johansen & 1.2231 & 0.8233 & 1.8172 & 0.8259 & 1.8114 &  1.0085 \\ 
$\tau_{max}$ &Probability transform incidence density CE& 1.2259 & 0.8294 & 1.8120 & 0.8317 & 1.8069 & 1.0075 \\ 
\cmidrule{3-8}
$\tau_{P90}$& Incidence proportion & 1.1882 & 0.7965 & 1.7724 & 0.8036 & 1.7567& 1.0240\\ 
$\tau_{P90}$& Probability transform incidence density  & 1.0594 & 0.8052 & 1.3938 & 0.7756 & 1.4469 &  0.8767\\ 
$\tau_{P90}$& 1-Kaplan-Meier & 0.9140 & 0.5808 & 1.4384 & 0.5939 & 1.4066 & 1.0553\\ 
$\tau_{P90}$ & Aalen-Johansen & 1.1882 & 0.7965 & 1.7724 & 0.8036 & 1.7567 &  1.0240 \\ 
$\tau_{P90}$ &Probability transform incidence density CE& 1.2062 & 0.8159 & 1.7833 & 0.8097 & 1.7970 & 0.9799 \\ 
\cmidrule{3-8}
$\tau_{P60}$ & Incidence proportion & 1.3867 & 0.8899 & 2.1608 & 0.9067 & 2.1207& 1.0468\\ 
$\tau_{P60}$ & Probability transform incidence density & 1.3719 & 0.9021 & 2.0863 & 0.9067 & 2.0759 & 1.0128  \\ 
$\tau_{P60}$ & 1-Kaplan-Meier & 1.3588 & 0.8683 & 2.1263 & 0.8883 & 2.0785 & 1.0569\\ 
$\tau_{P60}$ & Aalen-Johansen& 1.3867 & 0.8899 & 2.1608 & 0.9067 & 2.1207 & 1.0468 \\ 
$\tau_{P60}$ &Probability transform incidence density CE& 1.4800 & 0.9549 & 2.2937 & 0.9445 & 2.3191 & 0.9739 \\ 
\bottomrule
\end{tabular*}\hspace*{-2cm}
\end{table*}
\end{center}

The relative risks calculated by the incidence proportion and the Aalen-Johansen estimator only differ at the maximal follow-up time. This is a direct consequence of the last patient under observation in group B being censored leading to a slightly lower AE probability estimated by the incidence proportion than by the Aalen-Johansen estimator in group B. 

At all follow-up times, the relative risk estimated with the probability transform of the incidence density and the one estimated with the 1-Kaplan-Meier estimator are smaller than the relative risk obtained with the gold standard Aalen-Johansen estimator. At the maximum follow-up time, the direction of the estimated relative risks is different. But note that the 1 is always included in the confidence interval indicating no statistically significant therapy effects.

The relative risk estimated by the probability transform of the incidence density accounting for CEs and the relative risk estimated by the Aalen-Johansen are similar at the later follow-up times. At $\tau_{P60}$ the relative risk estimated by the probability transform of the incidence density accounting for CEs is greater than the one estimated by the Aalen-Johansen estimator.

The differences between the confidence intervals obtained with the model-based and with the bootstrapped variances are due to differences in the estimated variances. The possible impact of this is illustrated by the fact that the relative risk estimated with the Aalen-Johansen estimator is not included in the model-based confidence interval of the probability transform of the incidence density at the maximal follow-up time and at $\tau_{P90}$ but it is included in the confidence interval obtained with the bootstrapped variance.

Furthermore, the model-based confidence intervals of the incidence proportion and the Aalen-Johansen estimator are longer than the confidence interval where the variances are obtained with a bootstrap. For the probability transform of the incidence density and the 1-Kaplan-Meier estimator at $\tau_{max}^{(A)}$, $\tau_{max}^{(B)}$ and $\tau_{max}$, for the probability transform of the incidence density at $\tau_{P90}$ and for the probability transform of the incidence density accounting for CEs at $\tau_{P90}$ and $\tau_{P60}$ the confidence interval based on the bootstrapped variance is longer than the model-based confidence interval. This is also due to the differences in the estimated variances.

In summary, in this data example, estimators that lead to a higher estimate of the AE probability than the Aalen-Johansen estimator result in a smaller relative risk estimate than that based on the Aalen-Johansen estimator. The model based CIs of the probability transform of the incidence density may be too small in the sense that they do not cover the gold-standard estimate of the RR, but the bootstrapped CIs do not have this shortcoming.

\section{Simulation Study}\label{sec4}

In this section, we take a closer look at the three sources of bias namely, censoring, CEs and non-constant hazards. For this purpose, competing risk data with the event of interest and one CE are simulated for two independent groups A and B.\cite{beyersmann2011} Table \ref{tab3} describes the simulation scenarios. The simulation scenarios marked with $\star$ are displayed in the following. These represent all main findings of the investigation. The results of the other simulation scenarios can be found in the Supplementary Material. For each scenario $N=1000$ datasets are simulated.

\begin{center}
\begin{table*}[h!]%
\caption{Summary of the scenarios considered in the simulation study ($N=1000$).\label{tab3}}
\centering
\hspace*{-3cm}\begin{tabular*}{500pt}{@{\extracolsep\fill}lcccccc@{\extracolsep\fill}}
\toprule
\textbf{Scenario} & $\boldsymbol{\lambda_{\rm A}(t)}$  & $\boldsymbol{\bar\lambda_{\rm A}(t)}$ &$\boldsymbol{\lambda_{\rm B}(t)}$  & $\boldsymbol{\bar\lambda_{\rm B}(t)}$&$n_A=n_B$ & \textbf{censoring}\\
\midrule
S1 constant & 0.00265 & 0.00424 & 0.00246 & 0.00530& 200 & no \\{}\\
S2 constant $\star$ & 0.00265 & 0.00424 & 0.00246 & 0.00530& 400 & no \\{}\\
S3 constant $\star$ & 0.00265 & 0.00424 & 0.00246 & 0.00530 & 400 & 28\% in A; 15\% in B \\{}\\
S4 time-dependent & $\frac{1}{3}t^2$ & $\frac{8}{9}t$ & $\frac{1.8}{t+0.5}$& $\frac{8}{9}t$ & 400& no\\{}\\
S5 time-dependent  $\star$ & $\frac{1}{3}t^2$ & $\frac{8}{9}t$ & $\frac{1.8}{t+0.5}$& $\frac{8}{9}t$ & 400& 14\% in A; 10\% in B\\{}\\
S6 time-dependent & $\frac{1.8}{t+2}$ & $\frac{1}{2}t$ & $\frac{1.8}{t+2}$& $\frac{1}{8}t$ & 400& 18.5\% in A and B\\{}\\
S7 time-dependent & $\frac{1.8}{t+2}$ & $\frac{1}{2}t$ & $\frac{1.8}{t+2}$& $\frac{1}{8}t$ & 400& no\\{}\\
S8 time-dependent & $\frac{1}{2}t$& $\frac{1.8}{t+2}$ & $\frac{1}{8}t$& $\frac{1.8}{t+2}$ & 400& no\\{}\\
S9 time-dependent & $\frac{1}{2}t$& $\frac{1.8}{t+2}$ & $\frac{1}{8}t$& $\frac{1.8}{t+2}$ & 400& 18.5\%in A and B\\{}\\
S10 constant - time-dependent  $\star$ & 0.07 & $0.066\cdot t^{-0.283}$ & 0.06 &  $0.042\cdot t^{-0.283}$& 400 & 1.7\% in A; 2.3\% in B\\
\bottomrule
\end{tabular*}\hspace*{-2cm}
\end{table*}
\end{center}

Scenarios S1, S2 and S3 consider constant hazards for both events. The hazards are equal to the incidence densities estimated in the data example presented in Section~\ref{sec3}. The hazards $\lambda_{\rm A}(t)$ and $\lambda_{\rm B}(t)$ correspond to the AE hazards and the hazards $\bar\lambda_{\rm A}(t)$ and $\bar\lambda_{\rm B}(t)$ to the CE hazards in group A and group B, respectively. In contrast, scenarios S4 - S10 investigate the case where at least one hazard is time-dependent. In the scenarios S1, S2, S4, S7 and S8 the data is completely observed, i.e. without censoring. In scenarios S3, S5, S6 and S9 between 10 and 28 percent of the observations are censored whereas in the scenario S10 similar to the data example only very little censoring is present. The censoring times were generated from an uniform distribution and are independent of the event times.

\subsection{Probability estimators}\label{sec41}
Table~\ref{tab4} displays the simulation results with regard to the AE probability estimates. The table shows four important quantities of the comparison: the mean true simulated value, the mean Aalen-Johansen estimator which is considered the gold standard, the absolute mean bias with respect to the Aalen-Johansen estimator of the estimators and the relative mean bias with respect to the Aalen-Johansen estimator. 

For more stable results in the calculation of the mean true value and the mean Aalen-Johansen estimator the mean was calculated on the logit-transformed results and backtransformed to the probability scale. For the same reason in calculation of the relative mean bias with respect to the Aalen-Johansen estimator the mean was calculated on the log ratios and then exp-transformed back to the original scale. As in the data example the comparisons are also conducted at several follow-up times. Note that $\tau_{max}^{(A)}$, $\tau_{max}^{(B)}$, $\tau_{max}$, $\tau_{P90}$ and $\tau_{P60}$ take different absolute values in each simulation run. Therefore, as the true value depends on the follow-up time, it is not the same value for all simulation runs, but one for each simulation run. E.g., the true value for probability of the event of interest in group A is calculated by $P(T\le \tau,\epsilon=1\vert \text{group A})=\int\limits_0^\tau \exp(-\int\limits_0^u \lambda_{\rm A}(t)+\bar{\lambda}_{\rm A}(t)\text{d}t)\lambda_{\rm A}(u)\text{d}u$. For this reason, we calculate the mean true value over all $N=1000$ simulation runs. The reason why the other estimators are compared to the Aalen-Johansen estimator is two-fold. On the one hand, in the analyses of AE the true value is unknown. Therefore, the simulations show the bias induced by using one of the non-gold-standard estimators compared to using the gold-standard estimator. On the other hand, Table~\ref{tab4} also displays that the Aalen-Johansen value and the true value are equal rounding to the second decimal point.

The incidence proportion underestimates the AE probability if censoring is present, and therefore, at all follow-up times in scenarios S3, S5 and S10. Under a large percentage of censoring as in scenario S3 at the maximum follow-up time the AE probability estimated by the incidence proportion is on average about 24\% smaller than the AE probability estimated by the Aalen-Johansen estimator. As long as there is no censoring present the incidence proportion coincides with the Aalen-Johansen estimator (scenario S2 with a relative mean bias w.r.t. Aalen-Johansen of 0). 

\begin{center}
\begin{table*}[!h]
\caption{Simulation results: Mean Aalen-Johansen estimator (AJE), mean absolute and mean relative bias w.r.t. Aalen-Johansen estimator. The following follow-up times are considered $\tau_{max}^{(A)}$,$\tau_{max}^{(B)}$ the maximum follow-up times in group A and group B respectively, $\tau_{max}$ the minimum of the two maximal event times, $\tau_{P90}$ and $\tau_{P60}$ chosen according to the quantiles of the time and account for different lengths of follow-up between group A and group B. For a better display probability transform is abbreviated by PT. ($N=1000$)\label{tab4}}
\centering
\vspace{-0.3cm}
\footnotesize
\hspace*{-3cm}\begin{tabular*}{500pt}{@{\extracolsep\fill}clccccccccD{.}{.}{3}c@{\extracolsep\fill}}
\toprule
& & \multicolumn{4}{@{}c@{}}{\textbf{Group A} }& \multicolumn{4}{@{}c@{}}{\textbf{Group B} }\\
& & \multicolumn{4}{@{}c@{}}{\textbf{Scenario} }& \multicolumn{4}{@{}c@{}}{\textbf{Scenario} }\\
\cmidrule{3-6}\cmidrule{7-10}
\textbf{Follow-up time} & \textbf{Estimator}  &\textbf{S2}  & \textbf{S3}  & {\textbf{S5}}  & \textbf{S10}  &\textbf{S2}  & \textbf{S3}  & {\textbf{S5}}  & \textbf{S10}   \\
\midrule
& & \multicolumn{8}{@{}c@{}}{\textbf{mean true simulated values}}\\
\cmidrule{3-6}\cmidrule{7-10}
$\tau_{max}^{(A)}$,$\tau_{max}^{(B)}$ & true & 0.3837 & 0.3695 & 0.1305 & 0.6088 & 0.3163 & 0.3079 & 0.8551 & 0.6928 \\
  $\tau_{max}$ & true & 0.3826 & 0.3672 & 0.1219 & 0.6086 & 0.3161 & 0.3073 & 0.8551 & 0.6875 \\ 
  $\tau_{P90}$ & true & 0.3333 & 0.2990 & 0.0226 & 0.5329 & 0.2842 & 0.2586 & 0.8068 & 0.5571 \\ 
  $\tau_{P60}$ & true & 0.2128 & 0.1785 & 0.0010 & 0.3286 & 0.1891 & 0.1600 & 0.5708 & 0.3130 \\ 
\midrule
& & \multicolumn{8}{@{}c@{}}{\textbf{mean gold standard AE probability}}\\
\cmidrule{3-6}\cmidrule{7-10}
$\tau_{max}^{(A)}$, $\tau_{max}^{(B)}$&Aalen-Johansen & 0.3857 & 0.3717 & 0.1309 & 0.6107 & 0.3167 & 0.3077 & 0.8571 & 0.6925 \\ 
$\tau_{max}$ & Aalen-Johansen & 0.3840 & 0.3692 & 0.1195 & 0.6101 & 0.3162 & 0.3071 & 0.8571 & 0.6854 \\
$\tau_{P90}$ & Aalen-Johansen & 0.3335 & 0.3002 & 0.0191 & 0.5340 & 0.2846 & 0.2582 & 0.8042 & 0.5563 \\ 
$\tau_{P60}$ & Aalen-Johansen & 0.2128 & 0.1790 & 0.0003 & 0.3276 & 0.1888 & 0.1598 & 0.5640 & 0.3121 \\ 
& & \multicolumn{8}{@{}c@{}}{\textbf{Difference to AJE (mean absolute bias w.r.t. Aalen-Johansen)} }\\
\cmidrule{3-6}\cmidrule{7-10}
$\tau_{max}^{(A)}$, $\tau_{max}^{(B)}$ & Incidence proportion & -0.0000 & -0.0945 & -0.0223 & -0.0119 & -0.0000 & -0.0714 & -0.0275 & -0.0174 \\ 
$\tau_{max}^{(A)}$, $\tau_{max}^{(B)}$&PT incidence density  & 0.5259 & 0.3415 & 0.1364 & 0.3751 & 0.5508 & 0.3679 & 0.1400 & 0.2981 \\ 
$\tau_{max}^{(A)}$, $\tau_{max}^{(B)}$&1-Kaplan-Meier & 0.5466 & 0.3460 & 0.6155 & 0.3857 & 0.5683 & 0.3689 & 0.1074 & 0.3055 \\ 
$\tau_{max}^{(A)}$, $\tau_{max}^{(B)}$&PT incidence density CE & -0.0009 & -0.0007 & -0.0157 & -0.0021 & -0.0007 & -0.0007 & 0.0068 & -0.0025 \\ 
\cmidrule{3-6}\cmidrule{7-10}
$\tau_{max}$ & Incidence proportion& -0.0000 & -0.0922 & -0.0191 & -0.0119 & -0.0000 & -0.0709 & -0.0275 & -0.0164 \\ 
$\tau_{max}$  &PT incidence density & 0.4910 & 0.3283 & 0.0832 & 0.3748 & 0.5387 & 0.3612 & 0.1401 & 0.2879 \\
$\tau_{max}$  & 1-Kaplan-Meier   & 0.4973 & 0.3322 & 0.3328 & 0.3827 & 0.5458 & 0.3641 & 0.1069 & 0.2890 \\ 
$\tau_{max}$ &PT incidence density CE  & -0.0003 & -0.0005 & -0.0179 & -0.0017 & -0.0005 & -0.0008 & 0.0069 & 0.0005 \\ 
\cmidrule{3-6}\cmidrule{7-10}
$\tau_{P90}$ &Incidence proportion& -0.0000 & -0.0501 & -0.0015 & -0.0072 & -0.0000 & -0.0413 & -0.0204 & -0.0083 \\ 
$\tau_{P90}$&PT incidence density  & 0.2057 & 0.1398 & 0.0018 & 0.2285 & 0.2288 & 0.1561 & 0.0846 & 0.1537 \\ 
$\tau_{P90}$& 1-Kaplan-Meier & 0.2055 & 0.1392 & 0.0049 & 0.2309 & 0.2315 & 0.1564 & 0.0379 & 0.1538 \\ 
$\tau_{P90}$ &PT incidence density CE   & -0.0001 & -0.0002 & -0.0016 & 0.0082 & -0.0005 & -0.0004 & 0.0220 & 0.0086 \\ 
\cmidrule{3-6}\cmidrule{7-10}
$\tau_{P60}$&Incidence proportion  & -0.0000 & -0.0148 & -0.0000 & -0.0021 & -0.0000 & -0.0129 & -0.0072 & -0.0020 \\ 
$\tau_{P60}$ &PT incidence density & 0.0537 & 0.0348 & 0.0000 & 0.0734 & 0.0606 & 0.0395 & 0.0173 & 0.0451 \\ 
$\tau_{P60}$ & 1-Kaplan-Meier & 0.0532 & 0.0343 & 0.0000 & 0.0733 & 0.0607 & 0.0393 & 0.0053 & 0.0444 \\ 
$\tau_{P60}$ &PT incidence density CE & -0.0001 & -0.0001 & -0.0000 & 0.0063 & -0.0003 & -0.0003 & 0.0089 & 0.0046 \\ 
\midrule
& & \multicolumn{8}{@{}c@{}}{\textbf{Ratio to AJE - 1 (mean relative bias w.r.t. Aalen-Johansen)}}\\
\cmidrule{3-6}\cmidrule{7-10}
$\tau_{max}^{(A)}$, $\tau_{max}^{(B)}$&Incidence proportion& -0.0000 & -0.2535 & -0.1696 & -0.0196 & -0.0000 & -0.2312 & -0.0322 & -0.0251 \\ 
$\tau_{max}^{(A)}$, $\tau_{max}^{(B)}$&PT incidence density & 1.3654 & 0.9219 & 1.0341 & 0.6156 & 1.7405 & 1.1982 & 0.1638 & 0.4316 \\ 
$\tau_{max}^{(A)}$, $\tau_{max}^{(B)}$ & 1-Kaplan-Meier & 1.4145 & 0.9150 & 4.5261 & 0.6329 & 1.7813 & 1.1728 & 0.1254 & 0.4423 \\ 
$\tau_{max}^{(A)}$, $\tau_{max}^{(B)}$ &PT incidence density CE & -0.0024 & -0.0010 & -0.1189 & -0.0034 & -0.0022 & -0.0014 & 0.0079 & -0.0036 \\ 
\cmidrule{3-6}\cmidrule{7-10}
$\tau_{max}$&Incidence proportion & -0.0000 & -0.2492 & -0.1574 & -0.0195 & -0.0000 & -0.2300 & -0.0322 & -0.0240 \\ 
$\tau_{max}$&PT incidence density  & 1.2800 & 0.8917 & 0.6716 & 0.6157 & 1.7051 & 1.1786 & 0.1639 & 0.4210 \\  
$\tau_{max}$ & 1-Kaplan-Meier & 1.2910 & 0.8866 & 2.5790 & 0.6286 & 1.7137 & 1.1607 & 0.1247 & 0.4225 \\ 
$\tau_{max}$ &PT incidence density CE & -0.0008 & -0.0006 & -0.1492 & -0.0027 & -0.0015 & -0.0016 & 0.0081 & 0.0008 \\ 
\cmidrule{3-6}\cmidrule{7-10}
$\tau_{P90}$&Incidence proportion & -0.0000 & -0.1665 & -0.0713 & -0.0135 & -0.0000 & -0.1597 & -0.0254 & -0.0148 \\ 
$\tau_{P90}$&PT incidence density & 0.6165 & 0.4652 & 0.0811 & 0.4288 & 0.8048 & 0.6048 & 0.1054 & 0.2766 \\ 
$\tau_{P90}$ & 1-Kaplan-Meier & 0.6150 & 0.4617 & 0.2230 & 0.4334 & 0.8134 & 0.6040 & 0.0472 & 0.2765 \\ 
$\tau_{P90}$ &PT incidence density CE & -0.0004 & -0.0004 & -0.0746 & 0.0153 & -0.0019 & -0.0015 & 0.0273 & 0.0155 \\ 
\cmidrule{3-6}\cmidrule{7-10}
$\tau_{P60}$&Incidence proportion & -0.0000 & -0.0820 & -0.0254 & -0.0063 & -0.0000 & -0.0805 & -0.0127 & -0.0065 \\ 
$\tau_{P60}$&PT incidence density & 0.2513 & 0.1936 & 0.0034 & 0.2246 & 0.3212 & 0.2469 & 0.0307 & 0.1441 \\ 
$\tau_{P60}$ & 1-Kaplan-Meier & 0.2486 & 0.1902 & 0.0244 & 0.2245 & 0.3216 & 0.2451 & 0.0094 & 0.1417 \\ 
$\tau_{P60}$ &PT incidence density CE & -0.0003 & -0.0006 & -0.0159 & 0.0191 & -0.0018 & -0.0016 & 0.0158 & 0.0148 \\
\bottomrule
\end{tabular*}\hspace*{-2cm}
\end{table*}
\end{center}
\clearpage

The probability transform of the incidence density is always higher than the AE probability estimated with the Aalen-Johansen estimator. The 1-Kaplan-Meier estimator is also always higher. This difference is the most pronounced in scenario S5 in group A as this is the scenario with the most CEs. At the maximum follow-up time in group A the AE probability estimated by the 1-Kaplan-Meier estimator is on average about 452.61\% increased compared to the AE probability estimated by the gold standard. For smaller follow-up times the differences between the estimators are less pronounced since less CEs are present. Furthermore, it is notable that for group A in scenario S5 there is a huge difference between the estimate of the probability transform of the incidence density and the 1-Kaplan-Meier estimator. Due to the quadratic term in the hazard of the event of interest the assumption of constant hazards for the AE hazard is severely violated resulting in quite different estimates of the two estimators that censor CEs.

The bias between the probability transform of the incidence density accounting for CEs and the Aalen-Johansen estimator is negligible under constant hazards in scenario S2 and S3. On the other hand, if both hazards are non-constant (scenario S5) the probability transform of the incidence density accounting for CEs underestimates the event probability. In group B of scenario S5 at all follow-up times and in scenario S10 at $\tau_{P60}$ and $\tau_{P30}$ the probability transform of the incidence density accounting for CEs is higher than the Aalen-Johansen estimator. In all other scenarios respectively at all other follow-up times, it is the other way around.

The difference between the 1-Kaplan-Meier estimator and the Aalen-Johansen estimator is larger than the one between the parametric counterpart of the Aalen-Johansen estimator and the estimator itself. Only in scenario S5 in group B at $\tau_{P60}$ the mean bias w.r.t. the Aalen-Johansen estimator of the probability transform of the incidence density accounting for CEs is higher than the one of the Kaplan-Meier estimator. Generally, it seems that ignoring CEs may be more harmful than falsely assuming constant hazards.

\subsection{Variance estimators}\label{sec42}

\begin{figure}[h!]
\centerline{\includegraphics[width=400pt,height=500pt]{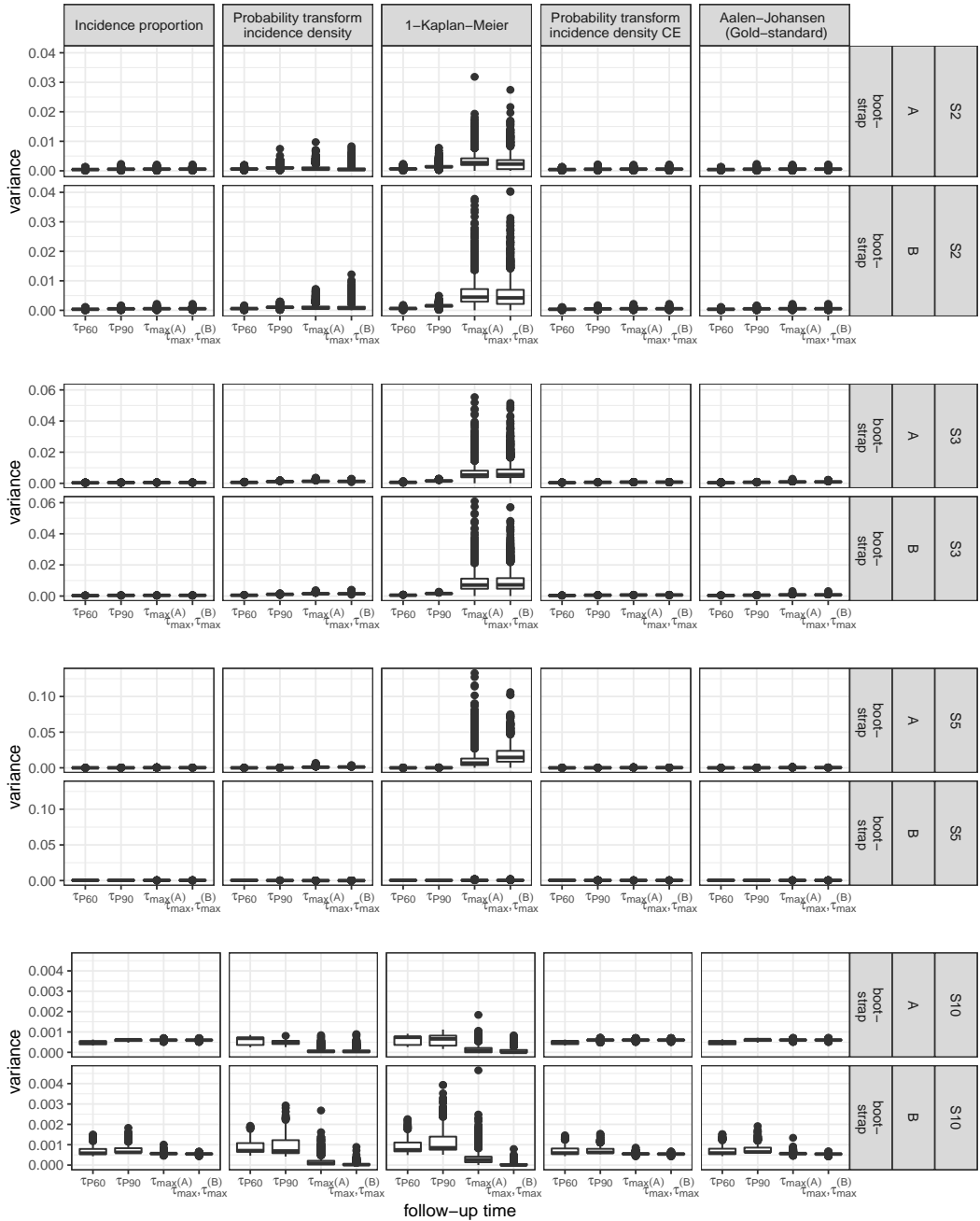}} 
\caption{Boxplots of the estimated variances for each simulation scenario. The variances are estimated by two different approaches, model-based and bootstrap. The following follow-up times are considered $\tau_{max}^{(A)}$,$\tau_{max}^{(B)}$ the maximum follow-up times in group A and group B respectively, $\tau_{max}$ the minimum of the two maximal event times, $\tau_{P90}$ and $\tau_{P60}$ chosen according to the quantiles of the time and account for different lengths of follow-up between group A and group B. ($N=1000$)\label{fig4}}
\end{figure}

Analogously to the data example the estimated variances are calculated model-based with the formulas from Section~\ref{sec32} and with a non-parametric bootstrap. The boxplots in Figure~\ref{fig4} display the estimated variances in all simulation scenarios for all considered probability estimators.

Considering the boxplots of the variances of the estimators one immediately notices the increased variances of the Kaplan-Meier estimator for the scenarios S2, S3 and for group A of S5. The reason for these outliers is that, if the last event is an AE, the Kaplan-Meier estimator equals 0 and as a consequence the 1-Kaplan-Meier estimate remaines unchanged at 1 near the end of follow-up in some of the bootstrap replications. The model-based variance using the Greenwood estimator is slightly smaller but still increased compared to the other estimators. This problem was already mentioned in Pocock et al. (2002)\cite{pocock2002} and is the motivation why we also consider some earlier follow-up times. Late censored observations and late CEs as present in group B of scenario S5 and in scenario S10 prevent the Kaplan-Meier estimator from dropping to 0 and therefore result in a smaller variance.

Furthermore, the parametric counterpart of the Kaplan-Meier estimator is less sensitive to the type of the last event. The variance of this estimator has only few outliers at the end of follow-up. In this comparison the parametric estimator has a smaller variance than the non-parametric Kaplan-Meier estimator.

In scenario S10 in group B an increased variance for earlier follow-up times can be detected. The reason is that censoring, increasing the variance, occurs early and the last event is rarely censored. 

However, it is notable that differences between the estimated variances of the non-parametric Aalen-Johansen estimator and its parametric counterpart are small. Furthermore, the bootstrapped variances of the incidence proportion, of the probability transform of the incidence density accounting for CEs and of the Aalen-Johansen estimator are comparable to the model-based ones. 

\subsection{Estimating the relative risk}\label{sec43}
In the data example, estimators that overestimate the AE probability underestimate the relative risk. This is further investigated in the simulations.

\begin{center}
\begin{table*}[t]%
\caption{Simulation results: true mean relative risk (RR) and mean relative risk calculated with the Aalen-Johansen estimator (AJE). The following follow-up times are considered $\tau_{max}^{(A)}$,$\tau_{max}^{(B)}$ the maximum follow-up times in group A and group B respectively, $\tau_{max}$ the minimum of the two maximal event times, $\tau_{P90}$ and $\tau_{P60}$ chosen according to the quantiles of the time and account for different lengths of follow-up between group A and group B. ($N=1000$) \label{tab5}}
\centering
\begin{tabular*}{500pt}{@{\extracolsep\fill}cccccD{.}{.}{3}c@{\extracolsep\fill}}
\toprule
&  \multicolumn{4}{@{}c@{}}{\textbf{Scenario} }\\
\textbf{Follow-up time} &\textbf{S2}  & \textbf{S3}  & {\textbf{S5}}  & \textbf{S10} \\
\cmidrule{2-5}
 & \multicolumn{4}{c}{\textbf{true simulated RR}}\\
\midrule
$\tau_{max}^{(A)}$,$\tau_{max}^{(B)}$ & 1.2132 & 1.2000 & 0.1526 & 0.8787 \\ 
  $\tau_{P60}$ & 1.2103 & 1.1950 & 0.1425 & 0.8853 \\ 
  $\tau_{P90}$ & 1.1728 & 1.1561 & 0.0280 & 0.9567 \\ 
  $\tau_{P60}$ & 1.1254 & 1.1157 & 0.0018 & 1.0499 \\ 
    & \multicolumn{4}{c}{\textbf{mean relative RR calculated with AJE}}\\
 \midrule
$\tau_{max}^{(A)}$,$\tau_{max}^{(B)}$ & 1.2180 & 1.2078 & 0.1527 & 0.8820 \\ 
  $\tau_{max}$ & 1.2142 & 1.2020 & 0.1394 & 0.8904 \\ 
  $\tau_{P90}$ & 1.1718 & 1.1625 & 0.0237 & 0.9605 \\ 
  $\tau_{P60}$ & 1.1267 & 1.1200 & 0.0004 & 1.0508 \\ 
\bottomrule
\end{tabular*}
\end{table*}
\end{center}

Table \ref{tab5} displays the true mean relative risk and mean relative risk where the probabilities were calculated by the Aalen-Johansen estimator. The relative risk calculated with the Aalen-Johansen estimator is comparable to the true relative risk. The different scenarios consider situations of a larger probability in group A (scenario S2, S3 and S10 at $\tau_{P60}$) and of a smaller probability in group A (scenario S5 and S10 except $\tau_{P60}$).

\begin{figure}[t]
\centerline{\includegraphics[width=400pt,height=340pt]{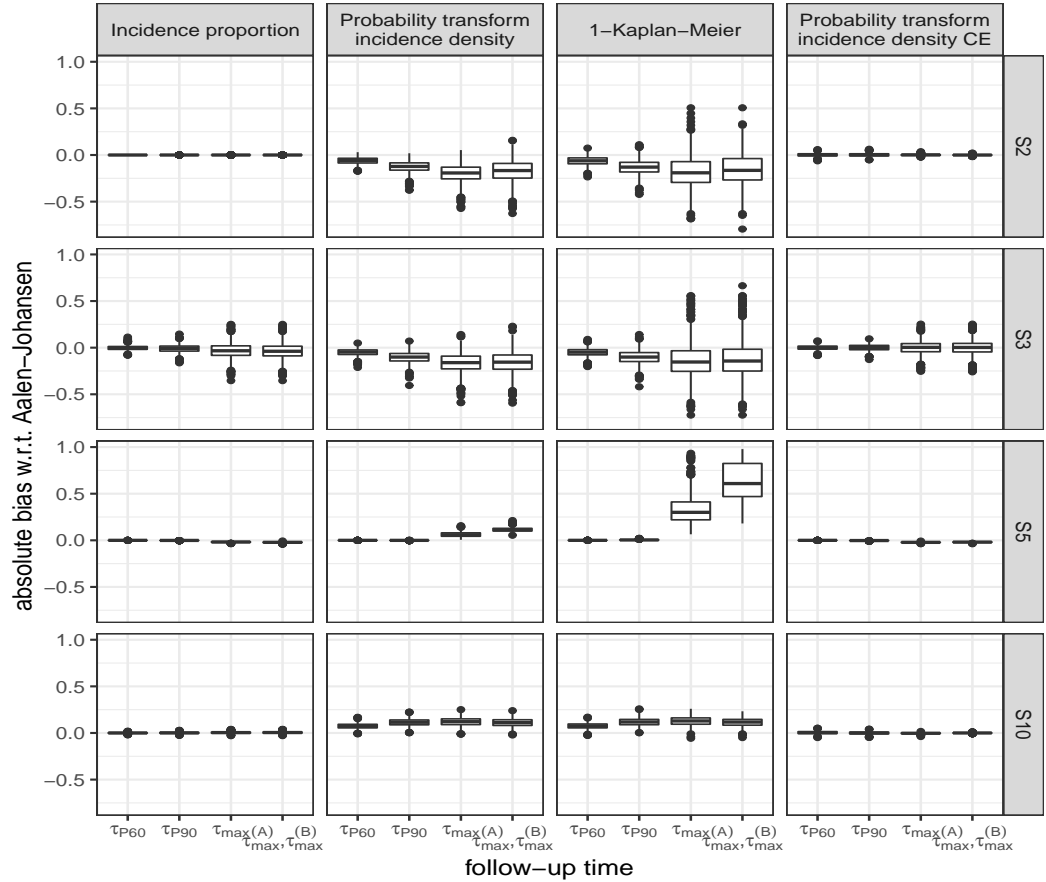}} 
\caption{Boxplots of the difference between the relative risk of the estimators of interest and the Aalen-Johansen estimator (absolute bias with respect to the Aalen-Johansen estimator). The following follow-up times are considered $\tau_{max}^{(A)}$,$\tau_{max}^{(B)}$ the maximum follow-up times in group A and group B respectively, $\tau_{max}$ the minimum of the two maximal event times, $\tau_{P90}$ and $\tau_{P60}$ chosen according to the quantiles of the time and account for different lengths of follow-up between group A and group B. ($N=1000$) \label{fig5}}
\end{figure}
\clearpage

\begin{center}
\begin{table*}[h!]%
\caption{Simulation results: Mean ratio of the relative risk of the estimator of interest and the relative risk calculated by the Aalen-Johansen estimator (AJE) - 1 (mean relative bias w.r.t. Aalen-Johansen). The following follow-up times are considered $\tau_{max}^{(A)}$,$\tau_{max}^{(B)}$ the maximum follow-up times in group A and group B respectively, $\tau_{max}$ the minimum of the two maximal event times, $\tau_{P90}$ and $\tau_{P60}$ chosen according to the quantiles of the time and account for different lengths of follow-up between group A and group B. For a better display probability transform is abbreviated by PT. ($N=1000$)\label{tab6}}
\centering
\begin{tabular*}{500pt}{@{\extracolsep\fill}cccccc@{\extracolsep\fill}}
\toprule
&&  \multicolumn{4}{@{}c@{}}{\textbf{Scenario} }\\\cmidrule{3-6}
\textbf{Follow-up time} & \textbf{Estimator}&\textbf{S2}  & \textbf{S3}  & {\textbf{S5}}  & \textbf{S10} \\
\midrule
$\tau_{max}^{(A)}$, $\tau_{max}^{(B)}$  & Incidence proportion& 0.0000 & -0.0290 & -0.1419 & 0.0057 \\ 
$\tau_{max}^{(A)}$, $\tau_{max}^{(B)}$  & PT incidence density & -0.1369 & -0.1257 & 0.7477 & 0.1286 \\ 
$\tau_{max}^{(A)}$, $\tau_{max}^{(B)}$  & 1-Kaplan-Meier  & -0.1319 & -0.1186 & 3.9105 & 0.1321 \\ 
$\tau_{max}^{(A)}$, $\tau_{max}^{(B)}$  & PT incidence density CE  & -0.0001 & 0.0004 & -0.1258 & 0.0002 \\ 
\midrule
$\tau_{max}$  &Incidence proportion & 0.0000 & -0.0249 & -0.1294 & 0.0047 \\
$\tau_{max}$  &PT incidence density  & -0.1571 & -0.1317 & 0.4363 & 0.1370 \\ 
$\tau_{max}$  &1-Kaplan-Meier & -0.1558 & -0.1269 & 2.1821 & 0.1449 \\ 
$\tau_{max}$  &PT incidence density CE & 0.0007 & 0.0010 & -0.1561 & -0.0035 \\ 
\midrule
$\tau_{P90}$ & Incidence proportion & 0.0000 & -0.0081 & -0.0471 & 0.0014 \\ 
$\tau_{P90}$ & PT incidence density  & -0.1043 & -0.0869 & -0.0220 & 0.1192 \\ 
$\tau_{P90}$ & 1-Kaplan-Meier & -0.1094 & -0.0887 & 0.1678 & 0.1229 \\
$\tau_{P90}$ & PT incidence density CE & 0.0015 & 0.0011 & -0.0992 & -0.0002 \\ 
\midrule
$\tau_{P60}$ & Incidence proportion & 0.0000 & -0.0017 & -0.0130 & 0.0002 \\ 
$\tau_{P60}$ & PT incidence density   & -0.0529 & -0.0428 & -0.0267 & 0.0704 \\ 
$\tau_{P60}$ & 1-Kaplan-Meier & -0.0553 & -0.0441 & 0.0149 & 0.0725 \\ 
$\tau_{P60}$ & PT incidence density CE  & 0.0015 & 0.0009 & -0.0315 & 0.0042 \\ 
\bottomrule
\end{tabular*}
\end{table*}
\end{center}

The boxplots in Figure~\ref{fig5} display the absolute mean bias with respect to the Aalen-Johansen estimator resulting from the other estimators. Using the incidence proportion or the probability transform of the incidence density accounting for CEs instead of the Aalen-Johansen estimator when comparing two treatment groups in terms of the relative risk induced no bias in the scenarios with constant hazards without censoring and time-dependent hazards. In the scenario with constant hazards and censoring (scenario S3) there are simulated datasets where the relative risk using the incidence proportion and probability transform of the incidence density accounting for CEs is either greater or smaller than the effect found by the Aalen-Johansen estimator. 

Using either the probability transform of the incidence density or  the 1-Kaplan-Meier estimator the relative risk is smaller than the relative risk calculated with the Aalen-Johansen estimator in scenario S2 and S3 where there is a beneficial effect of B but higher in scenario S5 at $\tau_{max}^{(A)}$, $\tau_{max}^{(B)}$ and $\tau_{max}$  and in scenario S10 at $\tau_{max}^{(A)}$, $\tau_{max}^{(B)}$, $\tau_{max}$ and $\tau_{P90}$ where a positive effect of A is simulated. But if there are severe differences between the two treatment groups A and B (scenario S5 at $\tau_{P90}$ and $\tau_{P60}$) using the probability transform of the incidence density may also result in a smaller relative risk than using the Aalen-Johansen estimator even though there is a beneficial effect of A observed.

The same conclusions can be drawn by considering the mean relative bias w.r.t. the Aalen-Johansen estimator of the relative risk (Table~\ref{tab6}). The relative risks calculated by the incidence proportion or the probability transform of the incidence density accounting for CEs show the most difference to the relative risk calculated by the Aalen-Johansen estimator in scenario S5. At the maximum follow-up time, the relative risk calculated by the incidence proportion is on average 14.15\% smaller than the relative risk calculated by the Aalen-Johansen estimator. The average relative ratio of the relative risk calculated with the probability transform of the incidence density and the relative risk calculated with the Aalen-Johansen estimator is between 4\% and 16\% in the scenarios S2 and S3. In scenario S5 at the maximum follow-up time the average mean ratio is the most pronounced. There the relative risk calculated by the Kaplan-Meier estimator is five times the size of the relative risk calculated by the gold-standard method.

\section{Conclusions and Discussion}\label{sec5} 
In this paper, we compared several estimators quantifying the AE probability. The gold-standard estimator of this probability in a time-to-event analysis with CEs is the Aalen-Johansen estimator. But simulations and the real data example illustrated that using the probability transform of the incidence density accounting for CEs or the incidence proportion may also provide unbiased estimates as long as the hazards are constant or there is no censoring at the considered follow-up time. The latter is usually not the case when adverse events are monitored during an ongoing trial by e.g. a data monitoring committee or a data safety monitoring board. These situations would be typically characterized by high amounts of censored observations. 

Often in AE analyses there are many CEs as progression, treatment discontinuation or death and only few observations may be censored. Therefore, the use of the incidence proportion may often be justified if censoring is low and the CEs have been specified appropriately. This emphasizes the need to consider thoroughly how the competing events are defined in the specific trial as this directly impacts the amount of censoring.
We do, however, recommend as a general rule to use one common $\tau$ when comparing incidence proportions, especially instead of the group specific maximal follow-up times if they differ. Moreover, the probability transform of the incidence density and the 1-Kaplan-Meier estimator always overestimate in the presence of CEs. A similar result was found when considering the relative risk. Using the 1-Kaplan-Meier estimator or the probability transform of the incidence density may result in the opposite conclusion about a therapy's safety as compared to using the Aalen-Johansen estimator. To summarize, in the literature \cite{kraemer2009,siddiqui2009,bender2019} mainly the constant hazards assumption is criticized, but one of our main results is that ignoring CEs and treating them falsely by censoring at the event time may be worse than misspecifying the model by falsely assuming constant hazards. This also calls for a larger emphasis on competing risk in the epidemiology literature on incidence densities. 

This paper emphazises the need to carefully consider the different follow-up times in safety analyses. One reason is that due to a small number of patients still at risk at the end of follow-up the variances of the estimators may be increased.\cite{pocock2002}  Another aspect with different follow-up times between groups is to calculate the relative risk using one common time point in both groups as otherwise the interpretation may be difficult.\cite{kunz2015} Here, we only considered the relative risk. But similar conclusions can be drawn for the risk difference (results not shown).

We further compared the model-based and the bootstrapped variance estimates of the estimators. Thereby, for the incidence proportion, the Kaplan-Meier estimator, the probability transform of the incidence density accounting for CEs and the Aalen-Johansen estimator no relevant differences between the two variance estimators were found. For the probability transform of the incidence density we found a smaller estimated variance for the model-based approach than for the bootstrapped one. The model-based variance assumes constant hazards whereas the non-parametric bootstrapped variance estimator does not rely on this assumption.\cite{hjort1992}

To summarize and to get back to the three questions this paper adds to the discussion about the analyses of AEs from the introduction:
The answer to question (i) is that the choice of the estimator is also crucial for group comparisons in terms of the relative risk. The bias in calculating the AE probabilities and variances of the AE probabilities do also directly influence the relative risk and the confidence intervals. Regarding question (ii), since the probability transform of the incidence density accounting for CEs is less biased with respect to the Aalen-Johansen estimator than of 1-Kaplan-Meier estimator, ignoring CEs is worse than falsely assuming constant hazards. When then considering question (iii), for earlier time points of analysis for most AE probability and relative risk estimators the bias with respect to the Aalen-Johansen estimator is smaller. Especially, for the incidence proportion the bias is almost negligible.

These analyses are motivated by the Survival analysis for AdVerse events with VarYing follow-up times project (SAVVY).\cite{stegherr2019} This is a joint project of academic institutions and pharmaceutical companies with the aim to improve standards for the reporting of incidences of adverse events. This project includes an empirical study calculating the above mentioned estimators in several randomized controlled trials and summarizing the results in a meta-analysis to assess the sources of bias in more real safety analyses.

AEs may also be recurrent\cite{siddiqui2009}, but the issue of CEs will remain as relevant as in time-to-first-event analyses. Recent papers by Charles-Nelson et~al.\cite{charlesnelson2019} and by Andersen et~al.\cite{andersen2019} emphazise the importance of considering competing (terminal) events also in analyses of recurrent events. See \cite{siddiqui2009,nelson2003,guettner2007} for suggestions on the analysis of recurrent AEs.

Moreover, here we focused on the comparison of probabilities and of the ratio of probabilities (relative risk) instead of the comparison of hazards. Interpretation of hazards and hazard ratios is often challenging in particular in the presence of CEs.\cite{beyersmann2007} The best way to communicate results is by visualizing the estimated probabilities for the two groups. All estimators introduced in Section~\ref{sec2} can easily be visualized in plots (see Figure~\ref{fig2}), although theses plots do not provide direct information about the possible dynamic pattern of the treatment effect.\cite{martinussen2018} Here, to better understand the differences between the estimated AE probabilities in the data example, we also considered a plot of the cumulative hazards estimated by the Nelson-Aalen estimator in Figure~\ref{fig3}. Therefore, as most analyses consider both the probabilities and the hazards one other point of future research and embedded in the SAVVY project is to compare different estimators of the hazard ratio, e.g., the hazard ratio obtained by the Cox proportional hazards model, the ratio of the Nelson-Aalen estimators and the ratio of the incidence density of both groups.

\section*{Acknowledgments}
RS was partially supported by Grant BE 4500/3-1 of the German Research Foundation (DFG). The DECIDER trial was supported by the German Federal Ministry of Education and Research (BMBF, Clinical Trials Program Grant 01KG0913).
%
%
%
\subsection*{Financial disclosure}

None reported.

\subsection*{Conflict of interest}

The authors declare no potential conflict of interests.

\subsection*{Data availability statement}
The data of the clinical trial are not publicly available due to confidentiality restrictions. The simulated data that support the findings of this study are available on request from the corresponding author.

%
%
%

\section*{Supporting information}

Additional Supporting Information may be found online in the supporting information tab for this article.

\appendix
\section{Proof: The estimand of the 1-Kaplan-Meier estimator is greater than the cumulative incidence function}\label{app1}
For the cumulative incidence function (left-hand side below) that can be estimated by the Aalen-Johansen estimator and the estimand of the 1-Kaplan-Meier estimator (right-hand side) the following inequality holds:
\begin{eqnarray*}
P(T\le t, \epsilon=1\vert \text{ group A})=\int\limits_0^t P(T\ge u-\vert \text{ group A})\lambda_{\rm A}(u)\text{d}u=\int\limits_0^t\exp\left(-\int\limits_0^u\lambda_{\rm A}(s)+\bar\lambda_{\rm A}(s)\text{d}s \right)\lambda_{\rm A}(u)\text{d}u
\\ \stackrel{\star}{\le}\int\limits_0^t\exp\left(-\int\limits_0^u\lambda_{\rm A}(s)\text{d}s\right)\lambda_{\rm A}(u)\text{d}u=1-\exp\left(-\int\limits_0^t\lambda_{\rm A}(u)\text{d}u\right)
\end{eqnarray*}

where relation $\star$ holds since $\exp\left(-\int\limits_0^u\bar\lambda_{\rm A}(s)\text{d}s\right)\le 1$ with equality if $\bar\lambda_{\rm A}(s)=0 \forall s$, i.e., no CEs are present in the data. Note that relation $\star$ does not postulate the existence of latent event-specific times. As a consequence, $\exp\left(-\int\limits_0^t\lambda_{\rm A}(s)\text{d}s\right)$ has no proper probability interpretation in settings with CEs.

\section{Model-based variances of the probability transform of the incidence density accounting for CEs}\label{app2}
It is known that 
\begin{equation*}
\sqrt{n}\left((\widehat{ID}_{\rm A},\widehat{\overline{ID}}_{\rm
        A})^T-(\theta_1,\theta_2)^T\right)\stackrel{d}{\rightarrow}Z\sim N\left(\begin{pmatrix}0\\0\end{pmatrix},\begin{pmatrix}{\rm var}(\theta_1)  & 0\\0 & {\rm var}(\theta_2)\end{pmatrix}\right).
\end{equation*}
and that the variance of the CE incidence density can be estimated analogously to the incidence density of the AE:
\begin{equation*}
{\rm var}(\widehat{\overline{ID}}_{\rm
        A})= \frac{\sum\limits_{u\in [0,\tau]} \bar{d}_{\rm A}(u)}{\left(\sum\limits_{i=1}^{n_{\rm A}}\min(t_i,\tau)\right)^2}.
\end{equation*}
The following mapping is defined to transform the two incidence densities to the estimator accounting for CEs on the probability scale:
\begin{equation*}
\Phi(x,y)=\frac{(1-\exp(-\tau(x+y)))\cdot x}{(x+y)}
\end{equation*}
Then the multivariate delta-method can be applied and the holds:
\begin{equation*}
\sqrt{n}(\Phi(\widehat{ID}_{\rm A},\widehat{\overline{ID}}_{\rm
        A})-\Phi(\theta_1,\theta_2))\stackrel{d}{\rightarrow}\Phi'(\theta_1,\theta_2) Z\sim N(0,s_{\rm A}^2)
\end{equation*}
where $s_{\rm A}^2$ is the variance of the probability transform of the incidence density and can be estimated by
\begin{eqnarray*}
\hat s_{\rm A}^2 =\left( \exp\left(-\tau\cdot\widehat{ID}_{\bullet \rm A}(\tau)\right)\cdot 
 \frac{\widehat{\overline{ID}}_{\rm
        A}(\tau)\cdot \left( \exp(\tau \cdot \widehat{ID}_{\bullet \rm A}(\tau))-1\right) +\tau \cdot \widehat{ID}_{\rm A}(\tau)\cdot \widehat{ID}_{\bullet \rm A}(\tau)}{\widehat{ID}_{\bullet \rm A}(\tau)^2} \right) ^2 \cdot\widehat{\rm var} (\widehat{ID}_{\rm A}(\tau)) \\
         +\left(\widehat{ID}_{\rm A}(\tau)\cdot \exp\left(-\tau\cdot\widehat{ID}_{\bullet \rm A}(\tau)\right)
         \cdot\frac{\tau \cdot \widehat{ID}_{\bullet \rm A}(\tau)-\exp\left(\tau \cdot \widehat{ID}_{\bullet \rm A}(\tau)\right)+1}{\widehat{ID}_{\bullet \rm A}(\tau)^2}\right)^2  \cdot \widehat{\rm var}(\widehat{\overline{ID}}_{\rm
        A}(\tau)).
\end{eqnarray*}
with $\widehat{ID}_{\bullet \rm A}(\tau)=\widehat{ID}_{\rm A}(\tau)+\widehat{\overline{ID}}_{\rm A}(\tau)$

\nocite{*}
\bibliography{AE}%

\clearpage
%
%

\end{document}